\documentclass[]{aa}  

\usepackage{graphicx}
\usepackage{txfonts}
\usepackage[]{hyperref}
\usepackage{color}
\usepackage{soul}

\begin{document} 

    \title{How well does surface magnetism represent deep Sun-like star dynamo action?}

   \titlerunning{Toroidal Flux Budget}
   \authorrunning{Finley et al.}

   \author{A. J. Finley\inst{1}
          \and
          A. S. Brun\inst{1}
          \and
          A. Strugarek\inst{1}
          \and
          R. Cameron\inst{2}
          }

    \institute{Department of Astrophysics-AIM, University of Paris-Saclay and University of Paris, CEA, CNRS, Gif-sur-Yvette Cedex 91191, France \\ \email{adam.finley@cea.fr} \and Max-Planck-Institut für Sonnensystemforschung, 37077 Göttingen, Germany}

   \date{Received March 30, 2023; accepted - -, -}

% \abstract{}{}{}{}{} 
% 5 {} token are mandatory

  \abstract{For Sun-like stars, the generation of toroidal magnetic field from poloidal magnetic field is an essential piece of the dynamo mechanism powering their magnetism. Previous authors have estimated the net toroidal flux generated in each hemisphere of the Sun by exploiting its conservative nature. This only requires observations of the photospheric magnetic field and surface differential rotation.}
  {We explore this approach using a 3D magnetohydrodynamic dynamo simulation of a cool
  star, for which the magnetic field and its generation is exactly known throughout the entire star.}
  {Changes to the net toroidal flux in each hemisphere were evaluated using a closed line integral bounding the cross-sectional area of each hemisphere, following the application of Stokes-theorem to the induction equation; the individual line segments corresponded to the stellar surface, base, equator, and rotation axis. The influence of the large-scale flows, the fluctuating flows, and magnetic diffusion to each of the line segments was evaluated, along with their depth-dependence.}
  {In the simulation, changes to the net toroidal flux via the surface line segment typically dominate the total line integral surrounding each hemisphere, with smaller contributions from the equator and rotation axis. The surface line integral is governed primarily by the large-scale flows, and the diffusive current; the later acting like a flux emergence term due to the use of an impenetrable upper boundary in the simulation. The bulk of the toroidal flux is generated deep inside the convection zone, with the surface observables capturing this due to the conservative nature of the net flux. }
  {Surface magnetism and rotation can be used to produce an estimate of the net toroidal flux generated in each hemisphere, allowing us to constrain the reservoir of magnetic flux for the next magnetic cycle. However, this methodology cannot identify the physical origin, nor the location, of the toroidal flux generation. In addition, not all dynamo mechanisms depend on the net toroidal field produced in each hemisphere, meaning this method may not be able to characterise every magnetic cycle.}

   \maketitle
%
%-------------------------------------------------------------------

\section{Introduction}

During the solar cycle \citep[see review of][and references therein]{hathaway2015solar}, the classical solar dynamo scenario pictures an originally poloidal magnetic field that is acted upon by the Sun's internal differential rotation transforming the field into toroidal wreaths at the base of the convection zone \citep[see review by][]{charbonneau2020dynamo}. Once this toroidal field becomes sufficiently strong, magnetic flux buoyantly rises towards the surface \citep{browning1984magnetic, birch2016low}, twisting in response to the Coriolis force \citep{d1993theoretical, fan2008three}, and convective motions \citep{jouve2012global}, producing Joy's law \citep{schunker2020average}. The net effect of which regenerates the poloidal field at the surface, but with the opposite polarity to the original poloidal field \citep{mordvinov2019evolution, noraz2022impact}. Although, the exact mechanism for regenerating the poloidal field is still debated, and could also depend on deep-seated cyclonic convection \citep{parker1955hydromagnetic, mason2002competition}, or magnetic buoyancy instabilities \citep{vasil2008magnetic}. An overview of these mechanisms can be found in \citet{brun2017magnetism}.

The evolution of the Sun's photospheric magnetic field has been continuously studied during the last four solar cycles (since the 1970s), from ground-based \citep{scherrer1977mean}, and later space-based \citep{scherrer1995solar, scherrer2012helioseismic}, observatories. Flux emergence begins at mid-latitudes and progresses down towards the equator during the approximately 11 year solar activity cycle \citep{van2015evolution}. The emerged magnetic field locally suppresses magnetoconvection at the surface, creating dark sunspots, and bright facular regions. The magnetic cycles of other Sun-like stars have begun to be mapped using magnetic activity proxies \citep{baliunas1985stellar, egeland2016evolution}, Zeeman-Doppler imaging \citep{saikia2016solar, saikia2018direct}, and the occultation of starspots by transiting exoplanets \citep[e.g.][]{morris2017starspots}. Although, not all cycles are observed to be solar-like \citep[see][]{jeffers2022crucial}

The regeneration of toroidal magnetic field from poloidal magnetic field remains poorly understood, despite the growing number of magnetic cycle observations. \citet{cameron2015crucial} proposed a method for evaluating the net toroidal flux in each hemisphere of the Sun that only requires observations of the photospheric magnetic field and the Sun's differential rotation pattern. The method assumes that the cancellation of toroidal flux between hemispheres can be treated as a simple loss term. However, this has not been explored in a scenario where the net toroidal flux in each hemisphere was actually known. In addition, the net toroidal flux may not scale with the total magnetic energy, due to cancellation effects, and thus fail to capture the underlying magnetic cycle.

If the net toroidal flux is important to the magnetic cycle of a star, this method has enormous value in estimating the reservoir of toroidal flux available for the next magnetic cycle. For the solar dynamo, some factors point to the net toroidal field being relevant including: the amount of net toroidal field estimated using the methodology of \citet{cameron2015crucial} being almost the same as is lost through flux emergence \citep{cameron2020loss};  changes in the amount of toroidal flux produced is reflected in the amount lost through emergence; and at times of maxima, flux emergence occurs with a systematic preference for the same orientation (at all latitudes in each hemisphere) in accordance with Hale’s law \citep{cameron2018observing}. However, this may not be the case for all dynamo mechanisms.

In this study, we apply the methodology from \citet{cameron2015crucial} to a cyclic dynamo simulation where the net toroidal flux in each hemisphere is known, and the fundamental assumptions of the method can be tested. Section 2 describes the methodology of \citet{cameron2015crucial}, and the magnetohydrodynamic simulation used in this work, published in \citet{brun2022powering}. Section 3 details the application of this method to the simulation, and observations of the Sun. The accuracy of this method in reproducing the deep internal stellar dynamo from surface measurements alone, its capability to diagnose the dynamo process, and the potential caveats, are summarised in Section 4.

\section{Methodology}

\begin{figure}
    \centering
    \includegraphics[trim=0cm 0cm 0cm 0cm, clip, width=0.5\textwidth]{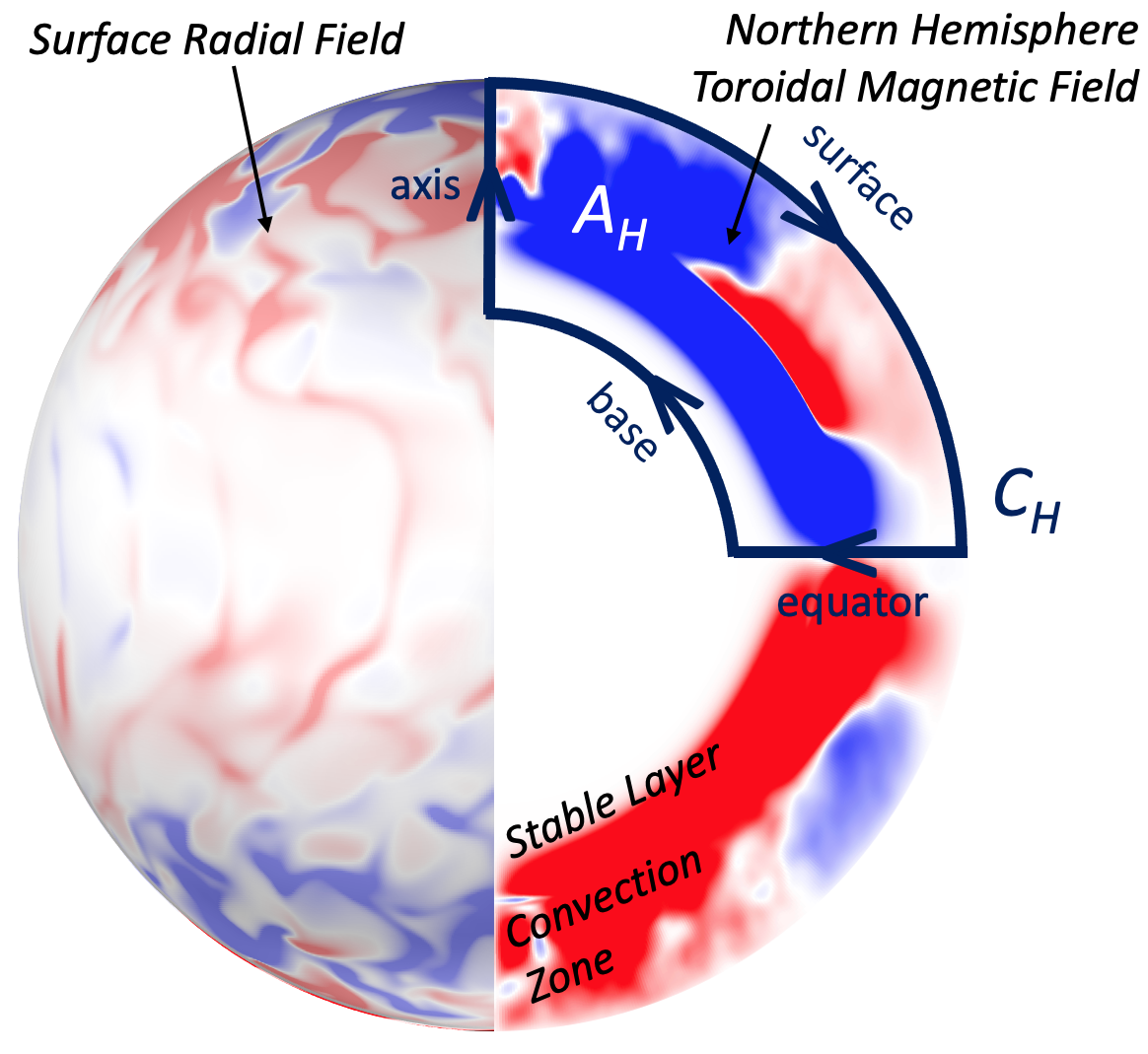}
    \caption{Schematic depiction of the Stokes-theorem contour, overlaid on a snapshot of the ASH simulation. The dark contour $C_H$ shows the path of integration around the area $A_H$ in the northern hemisphere, broken into four segments; surface, equator, base, and axis of rotation. The left hemisphere shows the structure in the radial field at the outer boundary of the simulation. The right hemisphere shows a slice-through the azimuthally averaged toroidal magnetic field in the simulation. }
    \label{fig:schematic}
\end{figure} 

\subsection{Evaluating the net toroidal flux generation}

The method proposed by \citet{cameron2015crucial} is based on the application of Stokes-theorem to the induction equation. First the induction equation is written in terms of the azimuthaly averaged large-scale magnetic field and velocity, $\bf{B}$ and $\bf{V}$, plus the turbulent magnetic field and velocity components, $\bf{b}$ and $\bf{v}$, such that,
\begin{equation}
    \frac{\partial \bf{B}}{\partial t} = \nabla \times  \bigg(\bf{V} \times \bf{B} + \langle \bf{v} \times \bf{b} \rangle - \eta \bf{J} \bigg),
\end{equation}
where $\eta$ is the magnetic diffusivity, $\bf{J}=$$1/\mu_0$$\nabla \times \bf{B}$ is the current density, and $\langle \bf{v} \times \bf{b} \rangle$ represents the azimuthaly averaged correlation between the turbulent magnetic field and velocity components.

Contours that enclose the northern and southern hemispheres are selected, denoted by $C_{H}$, each composed of four line integrals following the surface, base, equator, and axis of rotation (see dark contour in Figure \ref{fig:schematic}). By applying Stokes-theorem to the induction equation over the area inside the contour $A_{H}$, the time-evolution of the net toroidal flux in each hemisphere can be recovered using,
\begin{eqnarray}
    \frac{d\Phi_{H}}{d t} &=& \frac{d}{d t}\int_{A_{H}}B_{\phi}dA_{H}, \nonumber \\
     &=& \oint_{C_{H}} \bigg(\bf{V} \times \bf{B} + \langle \bf{v} \times \bf{b} \rangle - \eta \bf{J}\bigg) \cdot dl.
\label{eq:integral}
\end{eqnarray}
Each line segment contributes to the evolution of the net toroidal flux via the projection of the large-scale flows, the fluctuating flows, and magnetic diffusion onto the direction of the line segment. In spherical coordinates ($r,\theta,\phi$), for the northern hemisphere $A_{N}$, this is explicitly written,
\begin{eqnarray}
    \frac{d\Phi_{N}}{d t} &=& \int_\text{surface}^{0\text{ to }\pi/2} \bigg( V_{r}B_{\phi}-V_{\phi}B_{r} + \langle v_{r}b_{\phi}\rangle-\langle v_{\phi}b_{r} \rangle -\eta J_{\theta}\bigg)R_* d\theta \nonumber \\
                        &-& \int_\text{equator}^{R_*\text{ to }R_0} \bigg( V_{\theta}B_{\phi}-V_{\phi}B_{\theta} + \langle v_{\theta}b_{\phi}\rangle-\langle v_{\phi}b_{\theta} \rangle -\eta J_{r}\bigg)dr \nonumber \\
                        &-& \int_\text{base}^{\pi/2\text{ to }0} \bigg( V_{r}B_{\phi}-V_{\phi}B_{r} + \langle v_{r}b_{\phi}\rangle-\langle v_{\phi}b_{r} \rangle -\eta J_{\theta}\bigg)R_0 d\theta \nonumber \\
                        &+& \int_\text{axis}^{R_0\text{ to }R_*} \bigg( V_{\theta}B_{\phi}-V_{\phi}B_{\theta} + \langle v_{\theta}b_{\phi}\rangle-\langle v_{\phi}b_{\theta} \rangle -\eta J_{r}\bigg)dr,
\label{eq:integral_split}
\end{eqnarray}
where, $R_*$ and $R_0$ represent the radial distance to the surface and the base respectively. In this study, these line segments are evaluated individually, and represent the flux of toroidal field through each of these boundaries. Each line segment containing contributions from the large-scale flows, the fluctuating flows, and magnetic diffusion.

\begin{figure*}
    \centering
    \includegraphics[trim=0cm 0cm 0cm 0cm, clip, width=\textwidth]{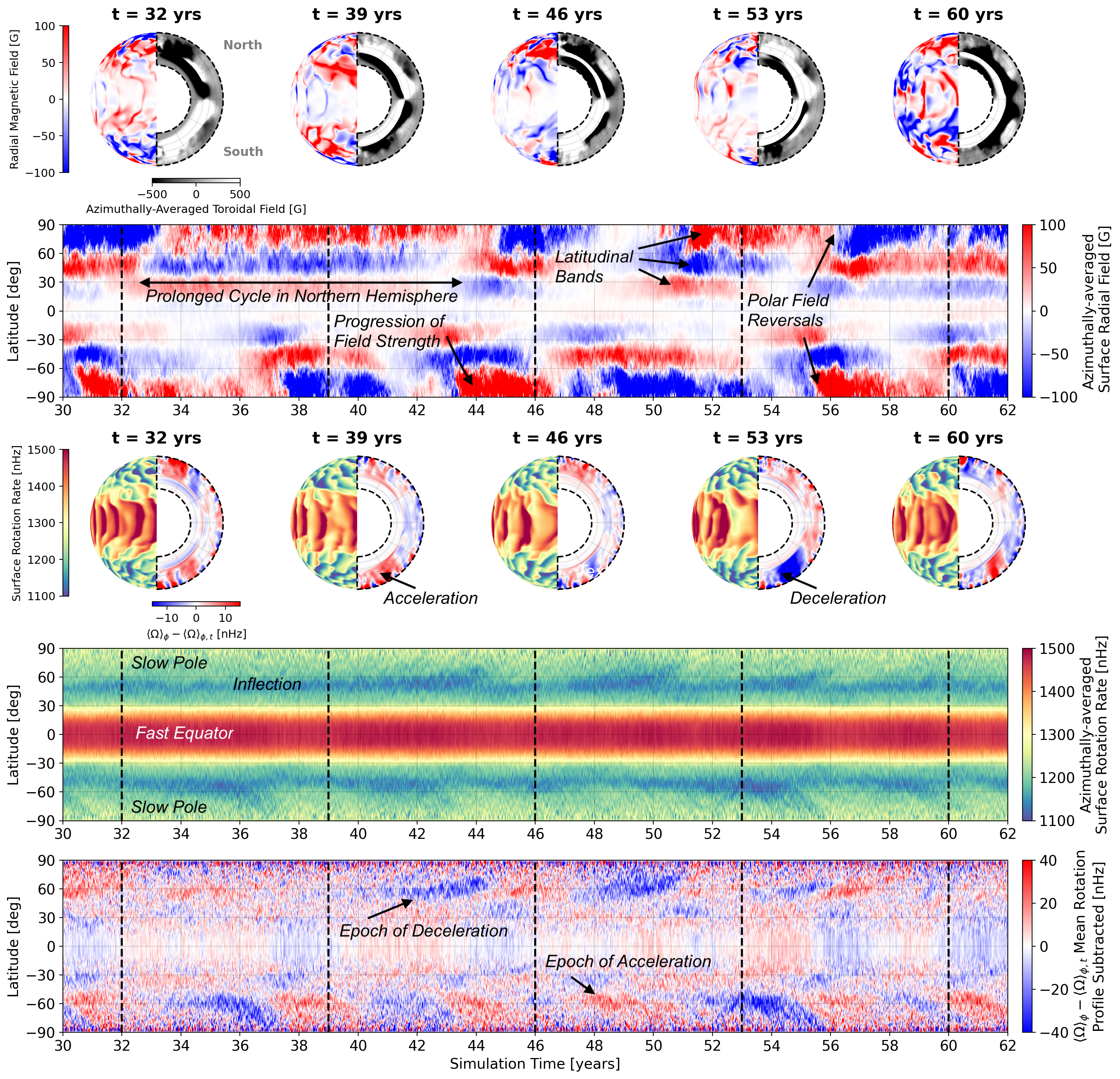}
    \caption{Overview of the ASH simulation used in this work. Snapshots of the simulation at five intervals are displayed in the top row, showing the radial field at the surface in the left hemisphere and the azimuthally averaged toroidal flux in the right hemisphere. The second panel shows the time-evolution of the azimuthally averaged radial magnetic field at the surface. The third row contains snapshots, similar to the top row, displaying the surface rotation rate in the left hemisphere and the azimuthally averaged rotation rate (minus the time-averaged profile) in the right hemisphere. The fourth panel shows the time-evolution of the azimuthally averaged surface rotation rate, with the fifth panel containing the residual rotation rate after the temporal average is subtracted (highlighting deviations coincident with the reversing of the radial magnetic field).}
    \label{fig:ash_example_M11_d3mag}
\end{figure*}

\begin{figure*}
    \centering
    \includegraphics[trim=0cm 0cm 0cm 0cm, clip, width=\textwidth]{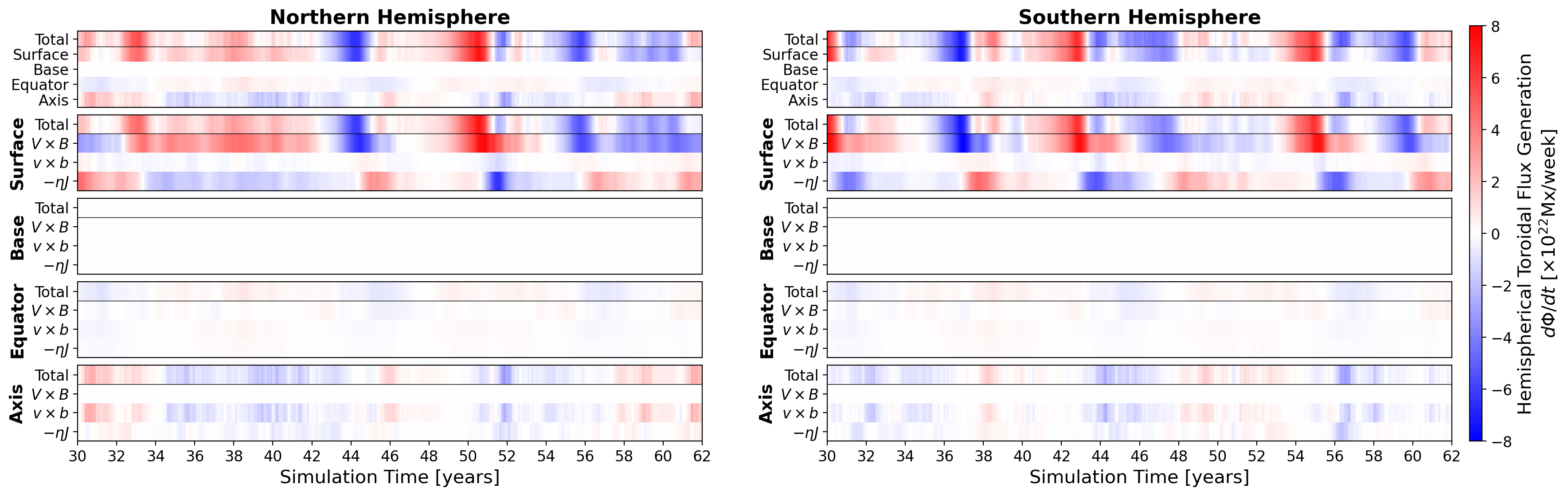}
    \caption{Breakdown of the net toroidal flux generation terms in the northern and southern hemispheres, left and right respectively. The top panel displays the contribution of each line integral (surface, base, equator, and axis) towards the total net toroidal flux generation in each hemisphere. The panels below, labelled by their corresponding line integral, further breakdown the contributions to each line integral in terms of the large-scale flows, the fluctuating flows, and magnetic diffusion. }
    \label{fig:model1_breakdown}
\end{figure*}

\subsection{Dynamo simulation}\label{simulations}

In this study, a global convective dynamo simulation performed using the ASH code \citep{brun2004global} was analysed. This simulation exhibits cyclic reversals of its large-scale magnetic field, with a solar-like differential rotation pattern. The star has a mass of 1.1$M_{\odot}$ and mean rotation rate three times higher than that of the Sun. The simulation domain is composed of a stable inner radiative zone, spanning 0.5 to 0.75 stellar radii, with an outer convective zone reaching up to 0.97 stellar radii. The simulation domain has a grid resolution of 769 radial ($r$) cells, 512 latitudinal ($\theta$) cells, and 1024 longitudinal ($\phi$) cells. Impenetrable and stress-free boundary conditions are maintained at the surface and base of the domain. The magnetic field is perfectly conducting at the base and matches a potential field at the surface. Convection is driven by a fixed entropy gradient at both boundaries. The ASH simulation (M11R3m) is described fully in \cite{brun2022powering} and references therein. 

The resulting cyclic reversals of the poloidal magnetic field have a periodicity of around five years. A temporal interval of 32 years, was selected for our analysis, spanning several magnetic cycles. Equation \ref{eq:integral_split} was computed at a weekly cadence throughout this period, i.e. on $\sim 1600$ ASH simulation snapshots. Figure \ref{fig:ash_example_M11_d3mag} contains snapshots of the simulation during this period, along with time-series of the surface magnetic field, and rotation rate, showing two full magnetic cycles (i.e. four polar field reversals). The polar field reversals in each hemisphere become desynchronised around $t=33$ years, and recover during the next few magnetic cycles. This indicates a strong contribution of the quadrupolar mode during this time \citep[see discussion in][]{derosa2012solar}. This ASH simulation possesses very interesting dynamo properties that
resemble that of a cyclic solar-like star. Though, it is by no means considered as being a perfect solar dynamo model. Here, we aim to characterise how representative surface observables are of the underlying cyclic dynamo, whose properties are similar to the Sun but not an exact representation of the Sun. At the base of the convection zone, the long-term cyclic magnetic field evolution hosts both equator-ward and polar branches (see Appendix \ref{branches}). This magnetic field is processed as it emerges through the convection zone to reach the surface. This produces a mostly dipole field that reverses on relatively long timescales. The increased mass and radius of the star creates some differences with respect to the solar dynamo. Here, the convection zone is shallower, the turnover timescale is shorter (around 24 days), and the rotation rate is three times larger, leading to a stellar Rossby number of 0.5 (slightly smaller than the Sun). The faster rotation also drives a differential rotation pattern with flows more concentrated at the equator. Accordingly, the azimuthally averaged surface magnetic field is structured into three latitudinal bands in each hemisphere. The central band in each hemisphere, having polarity opposite to the polar field, sits within an inflection of the surface differential rotation profile, where the latitudinal shear changes direction. These features have developed self-consistently in the simulation. When the surface radial fields reverse, the new magnetic field polarity begins to strengthen in the equator-ward band. The strengthening of the surface field then progresses towards the poles. Despite the opposing field polarity of the central band, the locally reversed latitudinal shear maintains a consistent sign of toroidal flux generation in each hemisphere. 

As the simulation's internal differential rotation shears the existing poloidal field into toroidal field, oppositely directed Maxwell stresses develops that resist the continued winding of the field. This can significantly quench the large-scale rotation with respect to the temporal average profile. Figure \ref{fig:ash_example_M11_d3mag} contains the azimuthally averaged surface rotation of the simulation in the fourth row, as well as the residual rotation rate after the temporal average is subtracted in the bottom row. Time periods of decreased rotation (coloured blue in the bottom panel) begin in the central bands and migrate over two to three years towards the poles. This kind of behaviour is somewhat similar to the torsional oscillation observed in the Sun \citep[see review of][]{howe2009solar}. During the period of desynchronisation in the northern hemisphere from $t=$ 33 to 45 years, the growth of toroidal field was insufficient to influence the internal differential rotation until $t=$ 41 years. The strength of the quenching at the surface is observed to vary from reversal to reversal. For example, in the southern hemisphere, the feedback of the reversals at $t=$ 43, 47, and 60 years was significantly smaller than the reversals at $t=$ 37 and 56 years (10nHz versus 30-40nHz decreases). Once the surface radial fields have reversed, the flow accelerates producing an enhancement with respect to the temporal average (coloured red in the bottom panel).

\section{Analysis of the toroidal flux budget}\label{results}

\subsection{Breakdown of each line segment}

\begin{figure*}
    \centering
    \includegraphics[trim=0cm 0cm 0cm 0cm, clip, width=\textwidth]{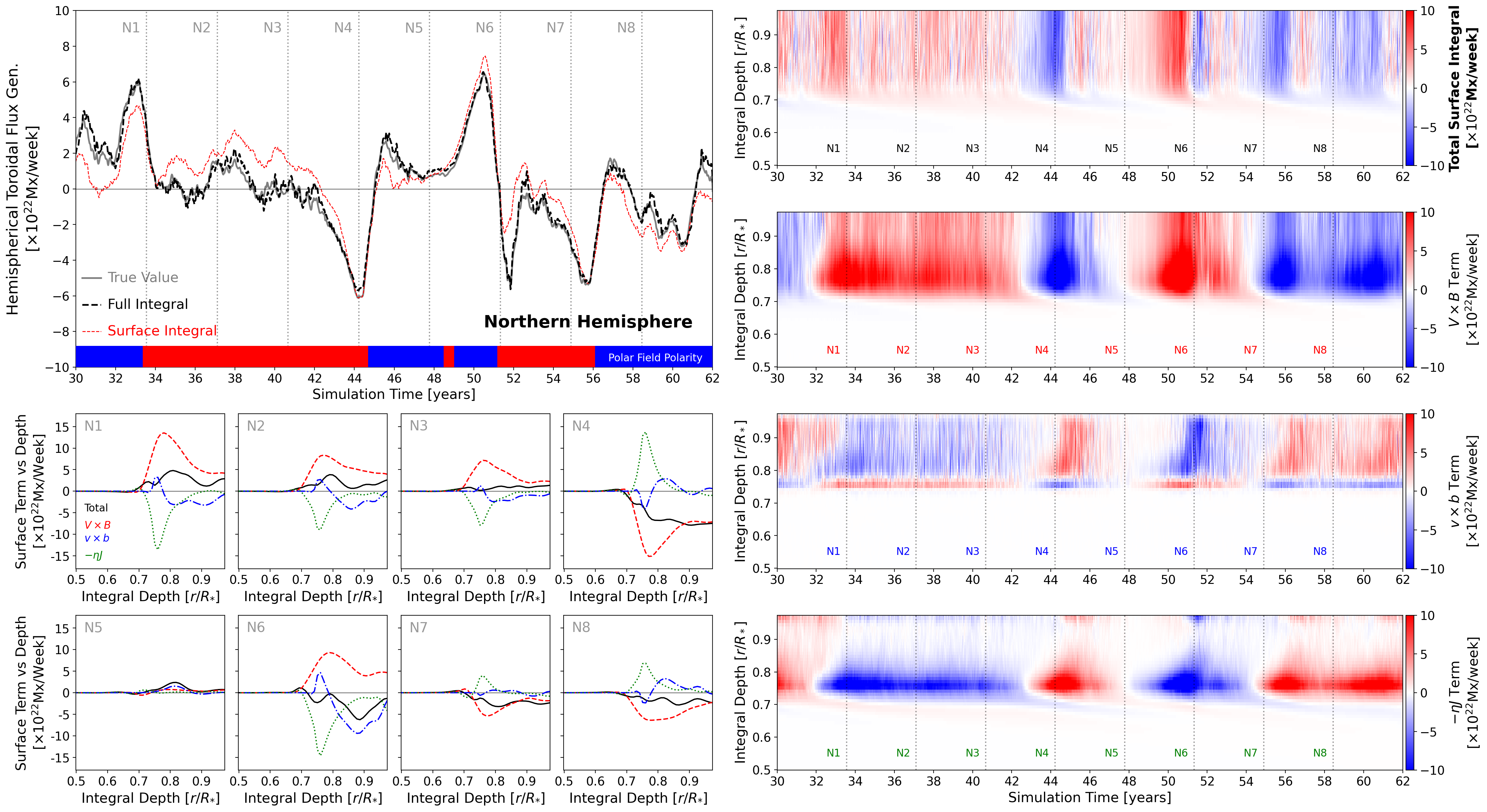}
    \caption{Comparison of the net toroidal flux generation in the northern hemisphere of the ASH simulation. The upper left panel shown the time-evolution of the net toroidal flux generation with the true value in solid grey, the full line integral in dashed black, and the surface line integral only in dotted red. The true value and line integral agree, with small differences due to the data sampling frequency (here data is taken at a weekly cadence rather than at the much smaller time-step cadence). Invervals labelled with vertical lines from one to eight, correspond to the smaller panels below. These panels show, for each interval, the depth dependence of the surface line integral, and its constituents (the large-scale flows, the fluctuating flows, and magnetic diffusion). The series of panels on the right contain a more continuous sampling of this parameter space, showing the time-evolving depth dependence of the surface line integral and its contributions. Above $\sim 0.75$ stellar radii, the line integral remains roughly constant with depth. Figure \ref{fig:southern} shows the same information but for the southern hemisphere.}
    \label{fig:northern}
\end{figure*}

The contribution from each individual line segment (surface, base, equator, and axis) surrounding the northern and southern hemispheres are displayed in Figure \ref{fig:model1_breakdown}. Values in Figure \ref{fig:model1_breakdown} (and later \ref{fig:northern}) have been smoothed, using boxcar-averaging with a six-month interval, in order to more easily make visual comparisons. The top panels show the net toroidal flux generation in each hemisphere, with the contributions from each of the four line segments, as a function of time. The surface line segment dominates the integral in each hemisphere, with smaller contributions from the equator and axis. The base line segment is located at the bottom of the ASH simulation domain, in the stable layer, and is therefore effectively zero. The lower four panels of Figure \ref{fig:model1_breakdown} show the net toroidal flux generation provided by each individual line segment, which is then further divided into the contributions from the large-scale flows, the fluctuating flows, and magnetic diffusion (see equation \ref{eq:integral_split}). For the surface line segment, the large-scale flows drive the overall sum, with magnetic diffusion having a smaller contribution that is typically in opposition. This is a result of diffusion acting as a means for the toroidal flux (built up by the large-scale flows) to emerge through the surface. As the boundary conditions of the ASH simulation require the radial velocity at the surface to tend to zero, this term acts like a replacement for flux emergence by advection. In other words, the flux of magnetic energy (or Poynting flux) is channelled through the other non-zeroed components \citep[see equations 4 to 7 in][]{finley2022stirring}. Hence, this is a loss term with respect to the net generation below. These observations hold for both the northern and the southern hemispheres.

The net toroidal flux generated in the northern hemisphere is plotted with a solid grey line in the top left panel of Figure \ref{fig:northern}. This was evaluated by performing a temporal derivative on the net hemispherical toroidal flux using the weekly simulation outputs, and smoothed to more easily make visual comparisons. This is compared with the resulting net toroidal flux generation calculated from the complete line integral with a dashed black line, and just the surface line segment with a dotted red line. The same information is available for the southern hemisphere in Appendix \ref{southern} Figure \ref{fig:southern}. The complete line integral matches the evolution of the net toroidal flux, with some small differences owing to the weekly cadence of the snapshots used in its calculation. The polarity of the northern polar magnetic field ($30^{\circ}$ around the pole) is displayed in red (positive) and blue (negative) along the bottom of the panel. The surface line integral appears to match best when the net toroidal flux generation is largest, for example during the reversals of the polar fields.  However, the surface line integral struggles to explain the net toroidal flux generation when the value is small and comparable with either the flux transported through the equator between hemispheres, or sources on the rotation axis. 

\subsection{Varying the depth of the surface line segment}

The toroidal flux in the ASH simulation is generated near the base of the convection zone \citep{brun2022powering}, far from the dynamics at the surface. This is also shown by the small panels in Figure \ref{fig:northern}, each one representing the calculation of the surface line segment at various depths in the northern hemisphere (in a sense, scanning through the simulation). The contributions from the large-scale flows, the fluctuating flows, and magnetic diffusion, are shown with coloured dashed, dash-dot, and dotted lines respectively. The total is then shown with a solid black line. Provided the other line segments closing the area below each surface line segment are small, this value represents the net toroidal flux generation in the ever diminishing area under the surface line segment. If the net toroidal flux generation inside the area remains the same, the value of the surface line segment should be constant. In general, this is the observed behaviour in the outer $10-20\%$ of the simulation, far away from the principal source of net toroidal flux. As the line segment moves to depths that contain imbalanced toroidal flux, due to the local generation of the opposing polarity, the value of the line segment begins to diverge. The largest values are found near the base of the convective zone. Notably, in the large-scale flows responsible for shearing the poloidal field into toroidal field, but also this is accompanied by a rise in magnetic diffusion that opposes this mechanism. 

The rightmost panels of Figure \ref{fig:northern} contain a more extensive breakdown of the depth-dependence of the surface line segment. The top panel showing the total surface integral, with the lower three panels breaking this down into the large-scale flows, the fluctuating flows, and magnetic diffusion. In the top panel, the line integral remains visibly constant down to around 0.85 stellar radii. This depth varies little in time, showing the dynamo action responsible for the cyclicality of the simulation is deeply-seated. Below 0.75 stellar radii, the integral passes into the radiative zone and so diminishes towards zero. The net toroidal flux generation, therefore lies in the lower half of the convection zone. The constituents of the line integral, have more structure with depth than their total. The large-scale flows, the principal driver of the net toroidal flux evolution, peak around 0.78 stellar radii. Magnetic diffusivity peaks slightly closer to the base of the convective zone, directly opposing the shearing from the large-scale flows. Further away from the base of the convection zone, the contribution from magnetic diffusion falls off, replaced by the turbulent flows which conserve the net toroidal flux generation up towards the surface. However, as the radial velocities are damped progressively towards zero in the outermost $\sim 5\%$ of the simulation, the turbulent flow contribution decreases near the surface with magnetic diffusion picking up this contribution.

\subsection{Comparison with the Sun}\label{results2}

\begin{figure*}
    \centering
    \includegraphics[trim=0cm 0cm 0cm 0cm, clip, width=\textwidth]{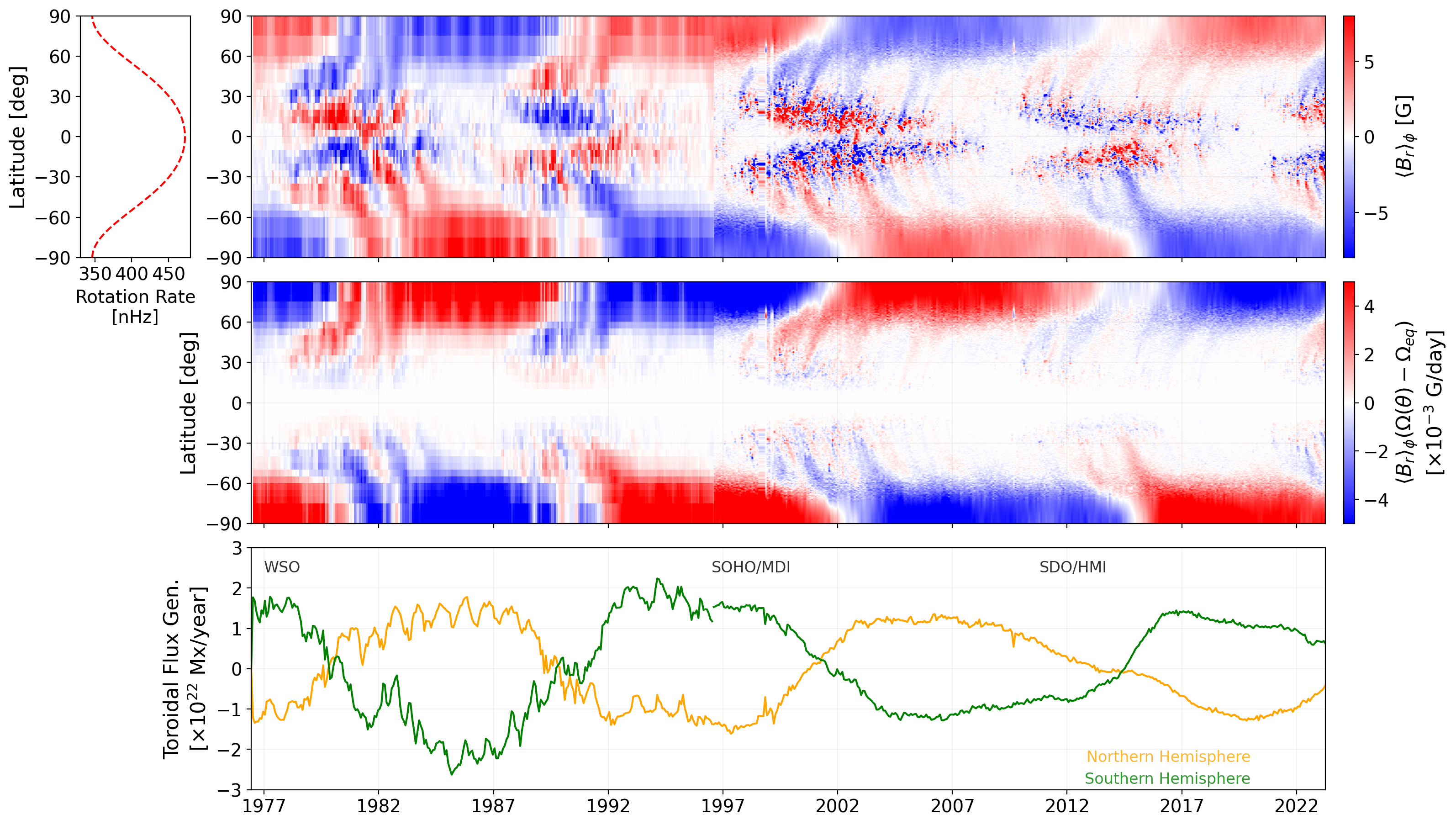}
    \caption{The method of \citet{cameron2015crucial} applied to magnetograms from WSO, SOHO/MDI, and SDO/HMI plus the surface rotation rate from \citet{snodgrass1983magnetic}. Top panels show the rotation profile and azimuthally averaged magnetic field. These are combined to produce the middle panel, a measure of the net toroidal field generated by the large-scale flows, from the surface observables. The lower panel shows the net toroidal flux generation in each hemisphere from this process.}
    \label{fig:cameron_real}
\end{figure*}

\begin{figure*}
    \centering
    \includegraphics[trim=0cm 0cm 0cm 0cm, clip, width=\textwidth]{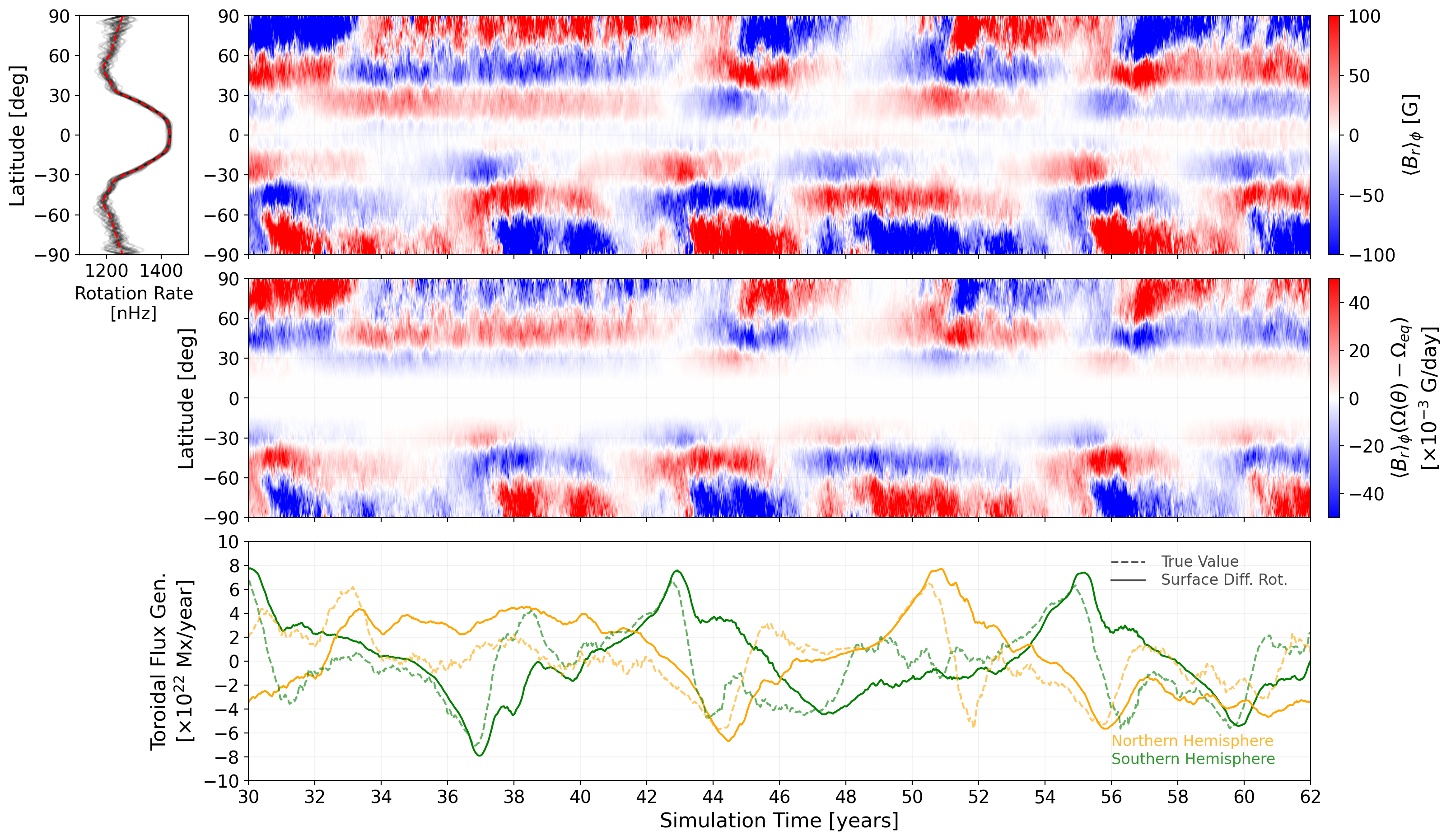}
    \caption{The method of \citet{cameron2015crucial} applied to the ASH simulation. Top panels show the rotation profile and azimuthally averaged magnetic field. These are combined to produce the middle panel, a measure of the net toroidal field generated by the large-scale flows, from the surface observables. The lower panel shows the net toroidal flux generation in each hemisphere from this process, in comparison to the true value from the simulation.}
    \label{fig:cameron_model}
\end{figure*}

 We now compare our analysis of the ASH simulation, to the same approach using observations from the Sun \citep[as in][]{cameron2015crucial}. Assuming that the dominant contribution to equation (\ref{eq:integral_split}) is the large-scale flows at the surface, then 
\begin{equation}
       \frac{d\Phi_{H}}{d t} = \int_{0}^{\pi/2} \bigg( V_{r}B_{\phi}-V_{\phi}B_{r} \bigg)R_* d\theta.
\end{equation}
If the primary contribution to this term is the large-scale differential rotation on the poloidal magnetic field $V_{\phi}B_{r}$. This can then be simplified to,
\begin{equation}
    \frac{d\Phi_{H}}{d t} = \int_{0^{\circ}}^{90^{\circ}} V_{\phi}B_{r} R_{*}d\theta = \int_{0}^{1} \bigg(\Omega(\theta)-\Omega_{eq}\bigg) B_{r} R_{*}^2 d\cos\theta,   
    \label{eq:cameron}
\end{equation}
in a rotating frame with the equator $\Omega_{eq}$, for the northern hemisphere \citep[see][]{cameron2015crucial}. Here, magnetograms are taken from the Wilcox Solar Observatory\footnote{Data accessed Jan. 2023: http://wso.stanford.edu/synopticl.html} (WSO, \citealp{scherrer1977mean}) and both the Michelson Doppler Imager (MDI, \citealp{scherrer1995solar}), onboard the Solar and Heliospheric Observatory (SOHO), and the Helioseismic and Magnetic Imager (HMI, \citealp{scherrer2012helioseismic}), onboard the Solar Dynamics Observatory (SDO). Both SOHO/MDI and SDO/HMI data products\footnote{Data accessed May 2023: http://hmi.stanford.edu/data/synoptic.html} include a polar field correction \citep{sun2011new,sun2018polar}. The radial magnetic field, shown in Figure \ref{fig:cameron_real}, is then multiplied by the solar surface differential rotation profile, taken form \citet{snodgrass1983magnetic},
\begin{equation}
    \Omega(\theta) = \Omega_{eq}+\alpha_2\cos^2\theta+\alpha_4\cos^4\theta,
    \label{omega_*}
\end{equation}
where $\Omega_{eq}=472.6$ nHz is the equatorial rotation rate, and the values of $\alpha_2=-73.9$ nHz and $\alpha_4=-52.1$ nHz. This forms the surface line segment integrand, shown in the second panel of Figure \ref{fig:cameron_real}, which is integrated in each hemisphere to estimate the net toroidal flux generation in the lower panel of Figure \ref{fig:cameron_real}. For the Sun, the integrand is dominated by the action of the differential rotation on the polar fields, which leads to the net toroidal flux generation in each hemisphere smoothly oscillating. This also implies that the peak net toroidal flux generation in each hemisphere occurs when the polar fields are largest.  

In comparison, the ASH simulation is more complicated to interpret. Figure \ref{fig:cameron_model} repeats this analysis for the simulation, only considering the surface differential rotation acting on the radial magnetic field. Compared to the Sun, the simulated differential rotation profile features a steeper profile around the lower latitude, and a flattening towards the poles. In addition, the modelled dynamo supports a latitudinally banded radial magnetic field, where opposite neighbouring polarities in each hemisphere add complexity to the interpretation. Nevertheless, the intergrand (second panel of Figure \ref{fig:cameron_model}) shares similarities with that of the Sun (in Figure \ref{fig:cameron_real}). The toroidal flux generation in each hemisphere peaking when the banded radial magnetic field strengthens and becomes the same polarity in the central and pole-ward bands. Then diminishes when these bands have mixed polarities. 

Figure \ref{fig:model1_breakdown} shows that the surface line segment of the simulation also includes a significant contribution from the magnetic diffusivity (essentially flux emergence). So equation (\ref{eq:cameron}) will not perform as well as the complete surface line segment, shown in contrast to the true value from the simulation in Figure \ref{fig:northern}. The loss of toroidal magnetic flux through the emergence of bipolar magnetic regions on the Sun is discussed in \citet{cameron2020loss} \citep[see also][]{jeffers2022crucial}. In Figure \ref{fig:cameron_model}, the true net toroidal flux production in the simulation is plotted with dashed lines for comparison with the estimate using only the differential rotation and radial magnetic field information. There are clearly significant departures from the true value, however these arise due to the strengthening of magnetic diffusivity. Therefore, the surface diagnostic captures the building of toroidal flux by the large-scale flows, up to its peak. After this, this method fails as the magnetic diffusivity increases, offsetting some of the toroidal flux being generated. The magnetic diffusivity of the simulation is indeed much larger than expected for the Sun, but in reality the emergence of bipolar magnetic regions may play a similar role. This should be accounted for in future estimations of the net toroidal flux production from surface observables alone.

\section{Conclusion}

The methodology presented in \citet{cameron2015crucial} allows us to estimate the net toroidal flux generation inside the Sun, using only the observed surface magnetic field and differential rotation pattern. This is a useful measure of the reservoir of toroidal flux that could emerge during the next magnetic cycle. However, this method has yet to be tested in a scenario where the net toroidal flux is actually known. In this study, we use a 3D magnetohydrodynamic dynamo simulation, where the net toroidal flux generation is known throughout the entire star, to determine the applicability and limitations of this methodology. This simulation has slightly different stellar parameters to the Sun, but hosts a cyclic dynamo and solar-like differential rotation. Following the method of \citet{cameron2015crucial}, the application of Stokes-theorem to the induction equation, allows us to evaluate the net toroidal flux generation from a closed line integral bounding the cross-sectional area of each hemisphere. The line segments of this integral correspond to the stellar surface, the base of the simulation domain, the equator, and the rotation axis. In order for this method to be useful, the surface segment should dominate the integral, as this is the only component that can be observed for the Sun and other stars.

Based on the application of this methodology to the dynamo simulation, we find that the surface segment is generally capable of describing the net toroidal flux generation. This is most reliable during the initial conversion of poloidal field into toroidal field during minima of activity, where the large-scale flows dominate the generation of toroidal flux. Although, to ensure the accuracy of the surface segment, the contributions of the turbulent flows and magnetic diffusion should also be included. Future studies of the Sun's toroidal flux generation should consider the contributions from the turbulent flows and magnetic diffusion \citep[see][]{cameron2020loss,jeffers2022crucial, 2023arXiv230502253C}. In the dynamo simulation, magnetic diffusion is large near the surface, acting as a pseudo flux emergence due to the impenetrable upper boundary. This component plays a significant role after the large-scale flows have acted to generate the bulk of the net toroidal flux. During phases of flux emergence, or maxima of activity, the toroidal flux generation is weak and so the equatorial and rotation axis segments can be comparable to the total integral. This introduces ambiguity in the net toroidal flux evolution predicted by surface observables alone during these periods. 

The success of this method, however,  does not imply that the surface differential rotation acting on the poloidal magnetic field is sufficient to generate the toroidal flux for the next magnetic cycle. It is clear that the toroidal flux is mostly generated, and confined, near to the base of the convective zone. Instead, the magnetic field lines are non-local and follow a conservation law. For this method to be applicable in real stars, the exchange of toroidal flux between hemispheres must be small or near zero in order for the unsigned toroidal flux to be a representative measure of the magnetic field being regenerated by the internal dynamo. Thus, if the toroidal flux is produced closer to the equator, with an increased possibility of flux transferring from one hemisphere to another, this method may begin to breakdown as the reservoir of toroidal energy in one hemisphere is not tightly linked to the surface flux evolution anymore. 

Equally, there are some dynamo mechanisms where the net toroidal flux in each hemisphere does not correlate with the magnetic activity cycle, as the net toroidal flux does not measure directly the total magnetic energy available to power the next cycle. One such example is the classic alpha-omega dynamo wave with a relatively high wave number. This dynamo produces several bands of toroidal flux of each polarity in each hemisphere. With appropriate boundary conditions there would be zero net flux in each hemisphere, with the bands cancelling, though substantial magnetic energy would be produced. Therefore, this method may not be applicable to every observed magnetic cycles.

\begin{acknowledgements}
This research has received funding from the European Research Council (ERC) under the European Union’s Horizon 2020 research and innovation programme (grant agreement No 810218 WHOLESUN), in addition to funding by the Centre National d'Etudes Spatiales (CNES) Solar Orbiter, and the Institut National des Sciences de l'Univers (INSU) via the Programme National Soleil-Terre (PNST).
Data manipulation was performed using the numpy \citep{2020NumPy-Array}, and scipy \citep{2020SciPy-NMeth} python packages.
Figures in this work are produced using the python package matplotlib \citep{hunter2007matplotlib}.
\end{acknowledgements}


\bibliographystyle{yahapj}
\bibliography{toroidalfluxgen}

\begin{appendix}

\section{Toroidal field evolution at the base of the convection zone}\label{branches}

\begin{figure*}
    \centering
    \includegraphics[trim=0cm 0cm 0cm 0cm, clip, width=\textwidth]{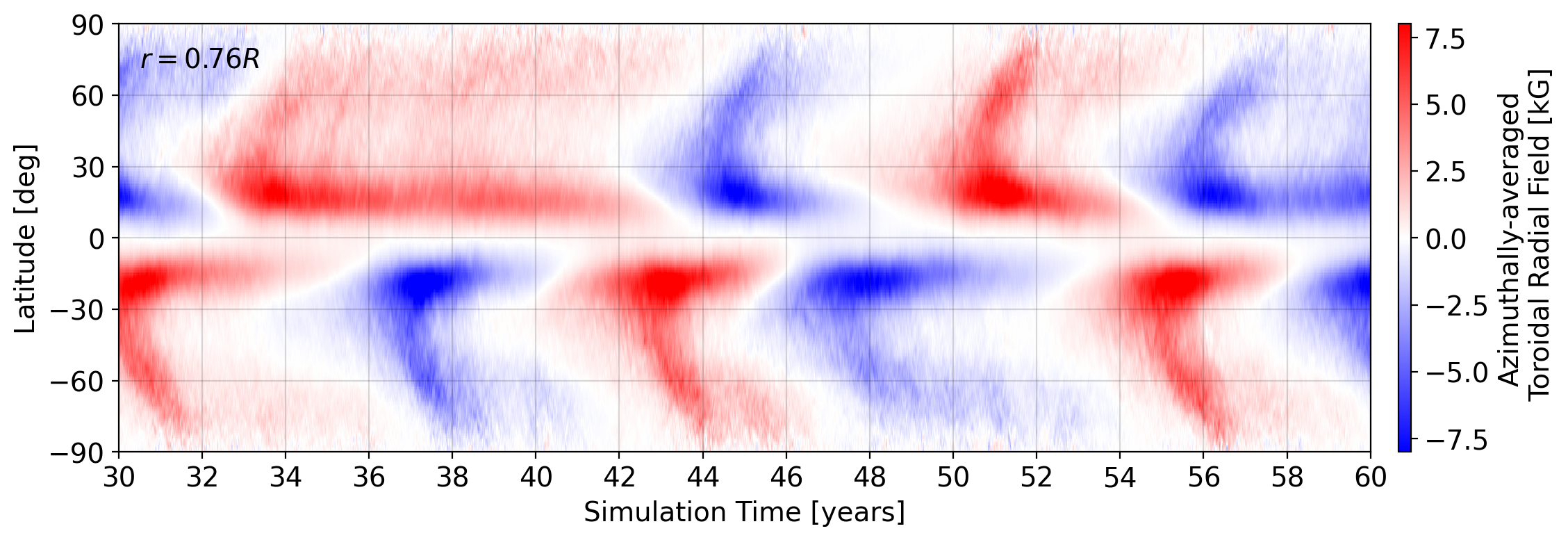}
    \caption{Time-evolution of the azimuthally averaged toroidal magnetic field at the base of the convection zone.}
    \label{fig:toroidalButterfly}
\end{figure*}

The ASH simulation investigated in this work (M11R3m) possesses some interesting dynamo properties that resemble a cyclic solar-like star \citep{brun2022powering}. This is best observed at the base of the convection zone, where the toroidal field is organised. Figure \ref{fig:toroidalButterfly} displays the azimuthally averaged toroidal field at a depth of 0.76 stellar radii. The equator-ward and polar branches are clearly visible, with the toroidal field in each hemisphere reversing after around five years.

\section{Analysis of southern hemisphere}\label{southern}

\begin{figure*}
    \centering
    \includegraphics[trim=0cm 0cm 0cm 0cm, clip, width=\textwidth]{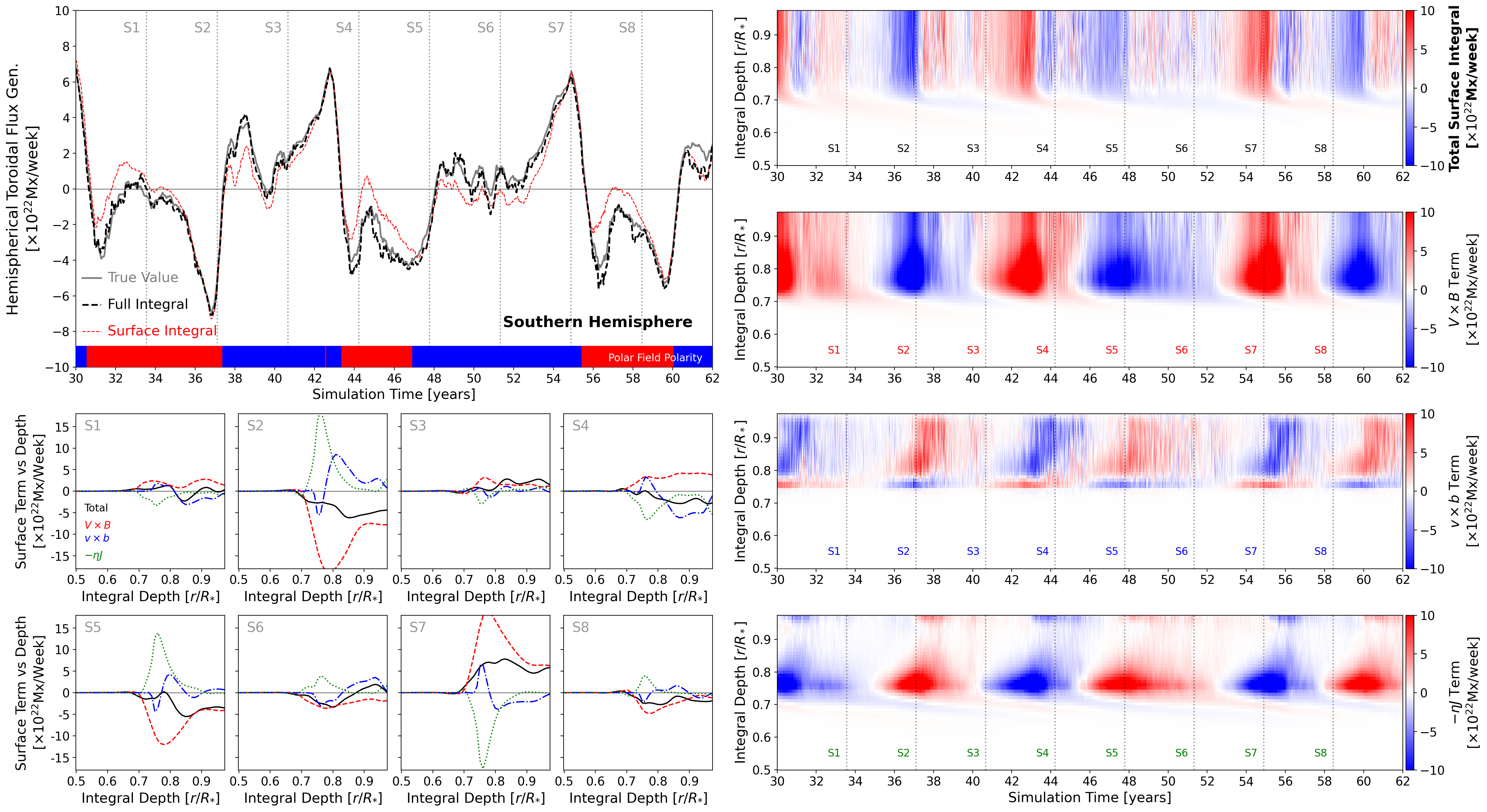}
    \caption{Same as Figure \ref{fig:northern}, but now for the southern hemisphere.}
    \label{fig:southern}
\end{figure*}

In Section 3.2, the \citet{cameron2015crucial} method for estimating the net toroidal flux generation, was compared against the true value in the ASH simulation for the northern hemisphere integral using equation (\ref{eq:integral_split}). Here the same analysis is performed on the southern hemisphere, where the integral is formed, 
\begin{eqnarray}
    \frac{d\Phi_{S}}{d t} &=& \frac{d}{d t}\int_{A_{S}}B_{\phi}dA_{S}, \nonumber \\
     &=& \oint_{C_{S}} \bigg(\bf{V} \times \bf{B} + \langle \bf{v} \times \bf{b} \rangle - \eta \bf{J}\bigg) \cdot dl.
\end{eqnarray}
and broken down into four line segments as,
\begin{eqnarray}
    \frac{d\Phi_{S}}{d t} &=& \int_\text{surface}^{0\text{ to }\pi/2} \bigg( V_{r}B_{\phi}-V_{\phi}B_{r} + \langle v_{r}b_{\phi}\rangle-\langle v_{\phi}b_{r} \rangle -\eta J_{\theta}\bigg)R_* d\theta \nonumber \\
                        &-& \int_\text{equator}^{R_*\text{ to }R_0} \bigg( V_{\theta}B_{\phi}-V_{\phi}B_{\theta} + \langle v_{\theta}b_{\phi}\rangle-\langle v_{\phi}b_{\theta} \rangle -\eta J_{r}\bigg)dr \nonumber \\
                        &-& \int_\text{base}^{\pi/2\text{ to }0} \bigg( V_{r}B_{\phi}-V_{\phi}B_{r} + \langle v_{r}b_{\phi}\rangle-\langle v_{\phi}b_{r} \rangle -\eta J_{\theta}\bigg)R_0 d\theta \nonumber \\
                        &+& \int_\text{axis}^{R_0\text{ to }R_*} \bigg( V_{\theta}B_{\phi}-V_{\phi}B_{\theta} + \langle v_{\theta}b_{\phi}\rangle-\langle v_{\phi}b_{\theta} \rangle -\eta J_{r}\bigg)dr.
\end{eqnarray}
The net toroidal flux generated in the southern hemisphere is plotted with a solid grey line in the top left panel of Figure \ref{fig:southern}, in comparison to the true value shown with a black dashed line, and the surface integral only with a dotted red line. The smaller panels in the lower right of Figure \ref{fig:southern} display radial scans through the simulation of the surface line integral. This is expanded in the right panels, showing the depth dependence of the net toroidal flux generation estimated from the surface line integral (split into the contributions from large-scale flows, fluctuating flows, and magnetic diffusion).

\end{appendix}

\end{document}